\documentclass[11pt,showkeys]{article}

\usepackage[dvipdfmx]{graphicx,xcolor}
\usepackage[fleqn]{amsmath}
\usepackage{newtxtext}%
\usepackage[varg]{newtxmath}%
\usepackage{latexsym}
\usepackage{enumerate}

\usepackage{bm}

\providecommand{\keywords}[1]{\textbf{\textit{Index terms---}} #1}
\newtheorem{remark}{Remark}

\title{Encoding and Decoding Algorithms of ANS Variants and Evaluation of Their 
Average Code Lengths\thanks{This paper is an English translation version of an invited paper
 published in the IEICE Transactions on Fundamentals of Electronics, Communications and Computer Sciences (Japanese Edition), DOI: 10.14923/transfunj.2024JAI0001, July 11, 2024.}\; \thanks{This paper is a reworked version of \cite{YI-STW2024} which was presented at the 12th Shannon Theory Workshop (STW2023).}}
\author{Hirosuke Yamamoto\thanks{The University of Tokyo, hirosuke@ieee.org.}
\and Ken-ich Iwata\thanks{University of Fukui, k-iwata@u-fukui.ac.jp.}}

\begin{document}

\maketitle
\begin{abstract}
Asymmetric Numeral Systems (ANS) proposed by Jarek Duda are high-performance distortionless data compression schemes that can achieve almost the same compression performance as arithmetic codes with less arithmetic operations than arithmetic coding.  
The ANS is widely used in various practical systems like Facebook, Apple, Google, Dropbox, Microsoft, and Pixar, due to their high performance, but many researchers still lack much knowledge about the ANS.
This paper thoroughly explains the encoding and decoding algorithms of the ANS, and theoretically analyzes the average code length achievable by the ANS.
\end{abstract}

\section{Introduction}
In conventional data compression coding like Huffman coding and arithmetic coding,
a data sequence $s^T=s_1s_2\cdots s_T$ is encoded and decoded in the
order of $s_t, t=1,2,\cdots, T$ \cite{YKAI}\cite{SG}.
But, Jarek Duda proposed Asymmetric Numeral Systems (ANS) to enhance Arithmetic coding, such that $s^T$ is encoded in backward order $s_t, t=T, \cdots, 2,1$, while
$s^T$ is decoded in forward order $s_t, t=1, 2, \cdots, T$ \cite{duda2008}--\cite{PDPCMM2023}.

Suppose that $c$ is the codeword obtained by arithmetic coding for a data sequence $s^T$.
Since the arithmetic code encodes $s^T$ in forward order of $s_t$, $t=1, 2, \cdots, T$,
it is determined by the order of the most significant bit (MSB) to the least significant bit (LSB) of the value of $c$.
When $s_{t}$ is encoded, the arithmetic encoder does not know the subsequent sequence $s_{t+1}^T=s_{t+1} \cdots s_T$.
Therefore, arithmetic codes are encoded using intervals of real numbers (or intervals of integers)
including $c$ so that they can handle any subsequent sequence $s_{t+1}^T$.
On the other hand, since the ANS encodes $s^T$ in backward order $s_t, t=T, T-1,\cdots, 1$,
the codeword $c$ is determined by the order of the LSB to the MSB. 
As a result, the ANS can encode and decode $s^T$
 using a single integer variable. This means that the ANS can achieve almost the same compression  rate as arithmetic codes with less arithmetic operations.
Due to this excellent feature, the ANS is utilized by Facebook Zstandard (ZSTD) compressor, Apple LZFSE compressor, Google, Dropbox, Microsoft, Pixar, etc.~\cite{PDPCMM2023}\cite{Wikipedia}\cite{HW2022}, and recently, many applied and related papers have been published \cite{HW2022}--\cite{YI-sita2023}.

However, many people are still unaware of ANS because many papers on ANS are only published on arXiv.org or as conference papers rather than as journal papers.
Furthermore, the algorithms and performance analyses in these papers are not written in an easy-to-understand manner, and there has been little information-theoretical evaluation. 
Therefore, even though people know the name of ANS, many of them do not know the detailed encoding-decoding algorithms and theoretical compression performance.

In this paper, we provide a detailed and easy-to-understand explanation of the ANS encoding and decoding algorithms, and present a new information-theoretical evaluation of the average code length that the ANS can achieve.

There are several variants of ANS. We treat ABS (Asymmetric Binary Systems) in Section 2, rANS (range variant of ANS) in Sections 3 and 4, and tANS (tabled variant of ANS\footnote{It is also called tabled ANS or table-based ANS.}) in Section 5.
In each section, we describe an encoding function, a decoding function, an encoding algorithm, and
a decoding algorithm, and demonstrate how these functions and algorithms can be used to encode and decode any data sequence correctly. 
Furthermore, we derive a strict information-theoretic upper bound on
the expected value of the average code length per source symbol, which we call 
 {\em the average code length} below for simplicity.
 
In this paper, we assume that 
a data sequence $s^T=s_1s_2\cdots s_T$, $s_t\in\mathcal{S}$, is
generated from an i.i.d.~source, which takes a value on
a finite discrete alphabet $\mathcal{S}$ with a probability distribution $p=\{p(s)\;|\; s\in\mathcal{S}\}$. For simplicity, the encoding and decoding algorithms are described assuming that
the probability distribution $p$ and sequence length $T$ are known.
We use the following notations\footnote{The notations used in this paper may be different from the original papers.}.
Let $\lg a=\log_2a$, and let $|\mathcal{A}|$ represent the cardinality of a set $\mathcal{A}$.
The entropy of the source is represented by $H(p)=\sum_{s\in\mathcal{S}}p(s)\lg (1/p(s))$,
and the relative entropy to a probability distribution $q=\{q(s)\;|\; s\in\mathcal{S}\}$
is denoted by $D(p\|q)=\sum_{s\in\mathcal{S}}p(s)\lg (p(s)/q(s))$.

\section{ABS}
In this section, we treat the case of binary source alphabet $\mathcal{S}=\{0,1\}$ with $p_1=p(1)$, $p_0=p(0)=1-p_1$, $0<p_1<1$.
In the encoding and decoding of the Asymmetric Binary Systems (ABS)
\footnote{ It is also called uABS (uniform ABS) \cite{duda2014}.}, a single integer variable $x$ is used. For simplicity, we assume that $x$ can have any number of digits. 
\vspace{0.1cm}
\subsection{Encoding and decoding procedures of ABS \cite{duda2008}--\cite{duda2014}}

\vspace{0.1cm}
\noindent \textbf{A. Definition of encoding function}\\
Encoding function $x_{t-1}:=C(s_t, x_{t})$ is defined by \eqref{eq2-1} and \eqref{eq2-2}.
When we want to specify whether $x_{t-1}$ is obtained by $s_t=0$ or $s_t=1$,
we represent it as $x_{t-1}=x_{t-1}^{(s_t)}$.

\vspace*{-0.2cm}
\begin{align}
x_{t-1}=x_{t-1}^{(0)}&:=\left\lceil\frac{x_t+1}{p_0}\right\rceil-1 \hspace{0.5cm}\text{if $s_t=0$},\label{eq2-1}\\
x_{t-1}=x_{t-1}^{(1)}&:=\left\lfloor \frac{x_t}{p_1}\right\rfloor\hspace{1.4cm}\text{if $s_t=1$}.\label{eq2-2}
\end{align}

\vspace{0.1cm}
\noindent \textbf{B. Definition of decoding function}\\
Decoding function $(s_t, x_t):=D(x_{t-1})$ is defined by \eqref{eq2-3}--\eqref{eq2-5}.
\begin{align}
s_t&:=\lceil (x_{t-1}+1)p_1\rceil-\lceil x_{t-1}p_1\rceil. \label{eq2-3}
\end{align}
For $x_{t-1}^{(s_t)}:=x_{t-1}$, 
\begin{align}
x_t&:=x_{t-1}^{(0)}-\lceil x_{t-1}^{(0)}p_1\rceil \hspace{0.5cm}\text{if $s_t=0$},\label{eq2-4}\\
x_t&:= \lceil x_{t-1}^{(1)}p_1\rceil \hspace{1.55cm}\text{if $s_t=1$}. \label{eq2-5}
\end{align}

\vspace{0.2cm}
\noindent \textbf{C. Encoding algorithm}
\vspace{-0.2cm}
\begin{enumerate}[a.]
 \setlength{\parskip}{0cm} 
 \setlength{\itemsep}{0cm} 
 \setlength{\itemindent}{0cm}
 \setlength{\labelsep}{0.2cm}
\item For a given data sequence $s^T=s_1s_2\cdots s_T$, set $x_T\leftarrow 1$.
\item Repeat $x_{t-1}\leftarrow C(s_t, x_{t})$ in backward order, $t=T, \cdots, 2, 1$. 
\item The codeword of $s^T$ is given by $x_0$.
\end{enumerate}

\noindent \textbf{D. Decoding algorithm}
\vspace{-0.2cm}
\begin{enumerate}[a.]
 \setlength{\parskip}{0cm}
 \setlength{\itemsep}{0cm}
 \setlength{\itemindent}{0cm}
 \setlength{\labelsep}{0.2cm}
\item Set a codeword $x_0$.
\item Repeat $(s_t, x_t)\leftarrow D(x_{t-1})$ in forward order, $t=1, 2, \cdots, T$.
\item $s^T=s_1s_2\cdots s_T$ is the decoded sequence.
\end{enumerate}

We prove that $D(x_{t-1})$ defined by \eqref{eq2-3}--\eqref{eq2-5} is the inverse function of 
$C(s_t, x_t)$ defined by \eqref{eq2-1}--\eqref{eq2-2}.
We first define $\hat{r}_t^{(0)}$ and $\hat{r}_t^{(1)}$ by
\begin{align}
\hat{r}_t^{(0)}&=x_{t-1}^{(0)}+1 -\frac{x_t+1}{p_0},\label{eq2-6}\\
\hat{r}_t^{(1)}&=\frac{x_t}{p_1}-x_{t-1}^{(1)}. \label{eq2-7}
\end{align}
From \eqref{eq2-1} and \eqref{eq2-2}, they satisfy $0\leq \hat{r}_t^{(0)}, \hat{r}_t^{(1)}<1$.
We next define $r^{(0)}_t$ and $r^{(1)}_t$ by $r^{(0)}_t=1-p_0(1-\hat{r}_t^{(0)})$ and $r^{(1)}_t=p_1\hat{r}_t^{(1)}$. Then, from relations $0<p_1=1-p_0\leq1-p_0(1-\hat{r}_t^{(0)})<1$ and
$0\leq p_1\hat{r}_t^{(1)}<p_1<1$, they satisfy
\begin{align}
0\leq r^{(1)}_t<p_1\leq r^{(0)}_t<1. \label{eq2-7-2}
\end{align}

From \eqref{eq2-6}, we have $x_t=x_{t-1}^{(0)}-[x_{t-1}^{(0)}p_1+\{1-p_0(1-\hat{r}_t^{(0)})\}]
=x_{t-1}^{(0)}-[x_{t-1}^{(0)}p_1+r^{(0)}_t]$.
Since  $x_t$ and $x_{t-1}^{(0)}$ are integers, $[x_{t-1}^{(0)}p_1+r^{(0)}_t]$ must be an integer.
Noting that $r^{(0)}_t$ satisfies \eqref{eq2-7-2}, we obtain $x_{t-1}^{(0)}p_1+r^{(0)}_t=\lceil x_{t-1}^{(0)}p_1\rceil$. Hence \eqref{eq2-4} holds.
On the other hand, from \eqref{eq2-7}, we have $x_t= x_{t-1}^{(1)}p_1+p_1\hat{r}_t^{(1)}= x_{t-1}^{(1)}p_1+r^{(1)}_t$. Since $x_t$ is an integer and $r^{(1)}_t$ satisfies \eqref{eq2-7-2},
we obtain $x_{t-1}^{(1)}p_1+r^{(1)}_t=\lceil x_{t-1}^{(1)}p_1\rceil$.
Hence, \eqref{eq2-5} holds.

From the above consideration, we have for any $s_t\in\{0,1\}$ that
\begin{align}
&\lceil x^{(s_t)}_{t-1}p_1\rceil =x^{(s_t)}_{t-1}p_1+r^{(s_t)}_t, \label{eq2-7-1}
\end{align}
and \eqref{eq2-3} can be derived as follows. 
\begin{align}
\lceil (x_{t-1}^{(s_t)}+1)p_1\rceil-\lceil x_{t-1}^{(s_t)}p_1\rceil 
&=\lceil (x_{t-1}^{(s_t)}+1)p_1-\lceil x_{t-1}^{(s_t)}p_1\rceil \rceil\nonumber\\
&= \lceil (x_{t-1}^{(s_t)}+1)p_1- (x_{t-1}^{(s_t)}p_1+r^{(s_t)}_t)\rceil\nonumber\\
&=\lceil p_1-r^{(s_t)}_t\rceil \nonumber\\
&=s_t, \label{eq2-8}
\end{align}
where the 2nd and 4th equalities hold from \eqref{eq2-7-1} and \eqref{eq2-7-2}, respectively.

\begin{remark}{\rm 
We can use \eqref{eq2-9} and \eqref{eq2-10}, instead of \eqref{eq2-1} and \eqref{eq2-2},
 and \eqref{eq2-10-1}--\eqref{eq2-10-3}, instead of \eqref{eq2-3}--\eqref{eq2-5}, 
 as the encoding function $C(s_t,x_t)$ and the decoding function $D(x_{t-1})$ \cite{duda2009}\cite{duda2014}.

\begin{align}
x_{t-1}&=x_{t-1}^{(0)}:=\left\lfloor \frac{x_t}{p_0}\right\rfloor\hspace{2.5cm}\text{if $s_t=0$},\label{eq2-9}\\
x_{t-1}&=x_{t-1}^{(1)}:=\left\lceil \frac{x_t+1}{p_1}\right\rceil-1\hspace{1.5cm}\text{if $s_t=1$},\label{eq2-10}\\
s_t&:=\lfloor (x_{t-1}+1)p_1\rfloor -\lfloor  x_{t-1}p_1\rfloor,  \label{eq2-10-1}\\
x_t&:=x_{t-1}^{(0)}-\lfloor  x_{t-1}^{(0)}p_1\rfloor  \hspace{2cm}\text{if $s_t=0$},\label{eq2-10-2}\\
x_t&:= \lfloor  x_{t-1}^{(1)}p_1\rfloor  \hspace{3.1cm}\text{if $s_t=1$}. \label{eq2-10-3}
\end{align}
}
\end{remark}

\subsection{Average code length of ABS} 
Substituting \eqref{eq2-4} and \eqref{eq2-5} into \eqref{eq2-7-1},
we obtain 
\begin{align}
\frac{x_{t-1}^{(0)}}{x_t}&=\frac{1}{p_0}\left(1+\frac{r_t^{(0)}}{x_t} \right),\label{eq2-11}\\
\frac{x_{t-1}^{(1)}}{x_t}&=\frac{1}{p_1}\left(1-\frac{r_t^{(1)}}{x_t}\right). \label{eq2-12}
\end{align}
Let $x_0(s^T)$ represent the codeword $x_0$ of a data sequence $s^T$.
Since $x_T=1$, the bit length\footnote{
Except for the MSB, the bit length of $x_0(s^T)$ is given by $\lfloor \lg x_0(s^T)\rfloor$ bits.
But, for simplicity, we use real number $\lg x_0(s^T)$ to represent the bit length of  $x_0(s^T)$ in this paper.} of $x_0(s^T)$ is given by $\lg x_0(s^T)=\sum_{t=1}^T \lg (x_{t-1}^{(s_t)}/x_t)$.
Therefore, $\lg (x_{t-1}^{(s_t)}/x_t)$ represents the increase of code length caused by encoding $s_t$.
We note from \eqref{eq2-7-2}, \eqref{eq2-11}, and \eqref{eq2-12}
that when $x_t$ is sufficiently large,  the relations $\lg (x_{t-1}^{(0)}/x_t)\approx-\lg p_0$ and $\lg (x_{t-1}^{(1)}/x_t)\approx -\lg p_1$ holds with very good accuracy.

We consider the case\footnote{It is not necessary to consider the case of $p_0=p_1=1/2$ because we cannot compress $s^T$ in this case. In the case of $p_0<p_1$, we use \eqref{eq2-9} and \eqref{eq2-10}.} of $0<p_1<1/2<p_0<1$.
We now define $\eta=\min\{1/p_0, 1/(2p_1)\}>1$. 
Then, from \eqref{eq2-11}, we have $x_{t-1}^{(0)}> x_t/p_0\geq \eta x_t$. 
Furthermore, noting $0\leq r_t^{(1)}<p_1<1/2$ and $x_t\geq 1$ in \eqref{eq2-12},
we also have $x_{t-1}^{(1)}> x_t/(2p_1)\geq \eta x_t$.
Hence, for any $t, 1\leq t \leq T$ and any $s\in\{0,1\}$,
$x_{t-1}^{(s)}$ satisfies $x_{t-1}^{(s)}> \eta x_t> \eta^{T-t+1} x_T=\eta^{T-t +1}$.
From this inequality, \eqref{eq2-7-2}, \eqref{eq2-11}, and \eqref{eq2-12},
$\lg (x_{t-1}^{(s)}/x_t)$ is bounded by
\begin{align}
\lg \frac{x_{t-1}^{(s)}}{x_t}&< \lg \frac{1}{p_s}\left(1+\frac{1}{x_t}\right)=\lg \frac{1}{p_s}+\lg \left(1+\frac{1}{x_t}\right)
\leq \lg \frac{1}{p_s} +\frac{\lg e}{x_t}<\lg \frac{1}{p_s}+\frac{\lg e}{\eta^{T-t}}.
\label{eq2-12-2}
\end{align}

Using \eqref{eq2-12-2}, we can derive an upper bound of the average code length $L$ for the case of $x_T=1$ as follows.
\begin{align}
L&=\frac{1}{T}\sum_{s^T\in\mathcal{S}^T}p(s^T)\lg x_0(s^T)\nonumber\\
&=\frac{1}{T}\sum_{s^T\in\mathcal{S}^T}p(s^T)\sum_{t=1}^T \lg \frac{x_{t-1}^{(s_t)}}{x_t} \nonumber
\end{align}
\begin{align}
&=\frac{1}{T}\sum_{t=1}^T \sum_{s_t\in\{0,1\}}p(s_t)\lg \frac{x_{t-1}^{(s_t)}}{x_t}\nonumber\\
&<\frac{1}{T}\sum_{t=1}^T\sum_{s\in\{0,1\}}p_s\left(\lg \frac{1}{p_s}+
\frac{\lg e}{\eta^{T-t}}\right)\nonumber\\
&= \sum_{s\in\{0,1\}}p_s\lg \frac{1}{p_s}+\frac{\lg e}{T}\sum_{t=1}^T\eta^{t-T}\nonumber\\
&< H(p)+\frac{\lg e}{T}\frac{\eta}{\eta-1}, \label{eq2-12-3}
\end{align}
where $H(p)$ is the entropy of the source. 
Since we know $L\geq H(p)$ from the source coding theorem for variable length coding \cite[Theorem 5.3.1]{YKAI}, it holds from \eqref{eq2-12-3} that $L\to H(p)$ as $T\to\infty$.

\section{rANS}
While the ABS discussed in the previous section is designed only for binary sources, 
the ANS  (Asymmetric Numeral Systems) is devised for general sources.
The ANS uses only integer arithmetic operations, although the ABS uses real numbers
$p_0$ and $p_1$. 
The ANS that uses integer ranges in encoding and decoding, like arithmetic range coding, is called rANS  (range variant of ANS). 

In this section, we assume that the integer variable $x$ can take any large number of bits. However, we will discuss
the case of restricting $x$ to an appropriate bit size in the next section.

\subsection{Encoding and decoding procedures of rANS \cite{duda2014}}
Assume that $N$ and $N_s, s\in\mathcal{S}$ are integers satisfying 
$N=\sum_{s\in\mathcal{S}}N_s$ and $N_s/N\approx p(s)$.
For each $s\in\mathcal{S}$, we define $d_{s}$
by 
\begin{align}
d_{s}=\sum_{\hat{s}\prec s}N_{\hat{s}}, \label{eq3-20}
\end{align}
where $\prec$ denotes an arbitrarily given total order on $\mathcal{S}$.

\vspace{0.1cm}
\noindent \textbf{A. Definition of encoding function}\\
Encoding function $x_{t-1}:=C(s_t, x_{t})$ is defined by\footnote{In this paper, residue operation 
``$a \bmod b$'' is represented as ``$\bmod(a,b)$'' in the same way as Duda's paper \cite{duda2014}.}
\begin{align}
x_{t-1}&:=N\left\lfloor \frac{x_t}{N_{s_t}}\right\rfloor +d_{s_t} + \bmod(x_t, N_{s_t}).\label{eq3-21}
\end{align}

\vspace{0.1cm}
\noindent \textbf{B. Definition of decoding function}\\
Decoding function $(s_t, x_t):=D(x_{t-1})$ is defined by \eqref{eq3-22} and \eqref{eq3-23}.
\begin{align}
s_t&:=\min\left\{s: \bmod(x_{t-1},N)<\sum_{i\preceq s} N_i\right\}, \label{eq3-22}\\
x_t&:= N_{s_t}\left\lfloor \frac{x_{t-1}}{N}\right\rfloor+\bmod(x_{t-1},N)-d_{s_t}. \label{eq3-23}
\end{align}

\newpage
\vspace{0.1cm}
\noindent \textbf{C. Encoding algorithm}
\vspace{-0.2cm}
\begin{enumerate}[a.]
 \setlength{\parskip}{0cm}
 \setlength{\itemsep}{0cm}
 \setlength{\itemindent}{0cm}
 \setlength{\labelsep}{0.2cm}
\item For a given data sequence $s^T=s_1s_2\cdots s_T$, set $x_T\leftarrow 1$.
\item Repeat $x_{t-1}\leftarrow C(s_t, x_{t})$ in reverse order, $t=T, \cdots, 2, 1$.
\item The codeword of  $s^T$ is given by $x_0$.
\end{enumerate}

\noindent \textbf{D. Decoding algorithm}
\vspace{-0.2cm}
\begin{enumerate}[a.]
 \setlength{\parskip}{0cm}
 \setlength{\itemsep}{0cm}
 \setlength{\itemindent}{0cm}
 \setlength{\labelsep}{0.2cm}
\item Set a codeword $x_0$.
\item Repeat $(s_t, x_t)\leftarrow D(x_{t-1})$ in forward order, $t=1, 2, \cdots, T$.
\item $s^T=s_1s_2\cdots s_T$ is the decoded sequence.
\end{enumerate}

\begin{figure}[t]
 \begin{center}
   \includegraphics[width=10cm]{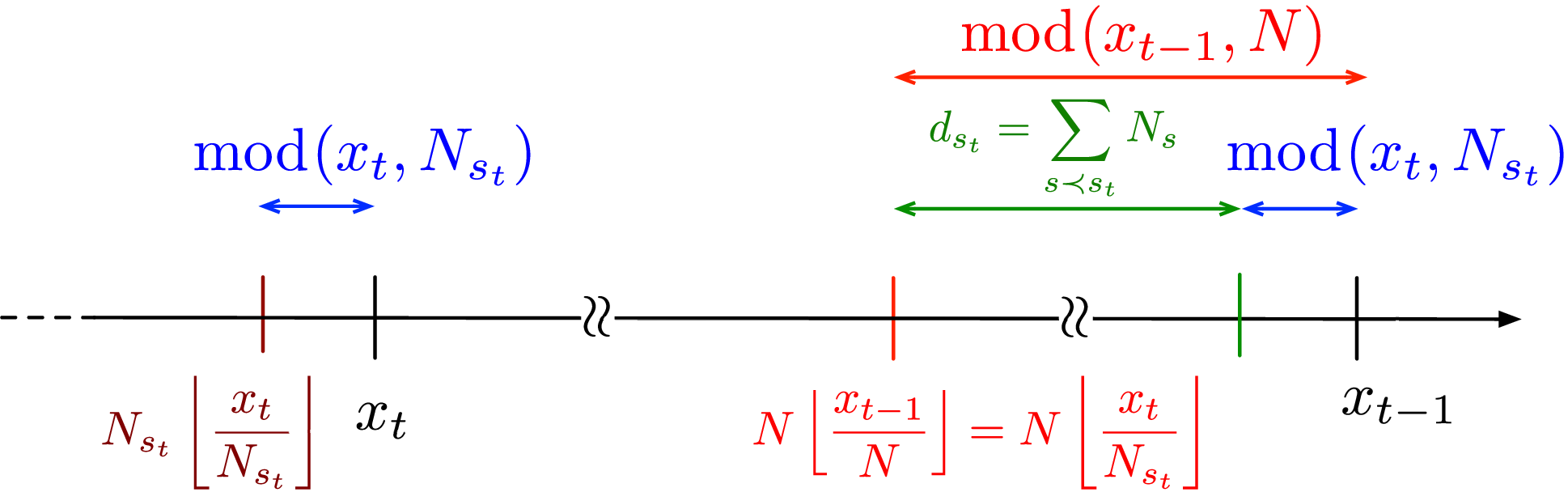}
    \caption{Relations used in rANS encoding and decoding.}
      \label{fig-rANS}
 \end{center}
\end{figure}

We now show that if $x_{t-1}$ is given by \eqref{eq3-21}, then $s_t$ and $x_t$ can be determined by \eqref{eq3-22} and \eqref{eq3-23}.

From \eqref{eq3-20} and $\bmod(x_{t-1}, N_{s_t})<N_{s_t}$, we obtain 
\begin{align}
d_{s_t}+\bmod(x_{t}, N_{s_t})<\sum_{s\prec s_t}N_{s}+N_{s_t}\leq N. \label{eq3-24}
\end{align}
Hence, the sum of the 2nd and 3rd terms on the right side of \eqref{eq3-21} is smaller than $N$.
By dividing both sides of \eqref{eq3-21} by $N$ and taking the floor function, we have
\begin{align}
\left\lfloor \frac{x_{t-1}}{N}\right\rfloor=\left\lfloor \frac{x_{t}}{N_{s_t}}\right\rfloor. \label{eq3-26}
\end{align}
Therefore, from \eqref{eq3-21}, \eqref{eq3-24}, and \eqref{eq3-26}, the following relation holds.
\begin{align}
\bmod(x_{t-1}, N)
&=x_{t-1}-N\left\lfloor \frac{x_{t-1}}{N}\right\rfloor\nonumber\\
&=x_{t-1}-N\left\lfloor \frac{x_{t}}{N_{s_t}}\right\rfloor\nonumber\\
&=d_{s_t} + \bmod(x_t, N_{s_t}) \label{eq3-24-2}\\
&<\sum_{s\prec s_t}N_{s}+N_{s_t}. \label{eq3-24-3}
\end{align}

Inequality \eqref{eq3-24-3} means that $s_t$ can be decoded by \eqref{eq3-22}.
On the other hand, \eqref{eq3-23} can be derived from \eqref{eq3-26} and \eqref{eq3-24-2}
as follows.
\begin{align}
x_t &=N_{s_t}\left\lfloor \frac{x_{t}}{N_{s_t}}\right\rfloor+\bmod(x_{t}, N_{s_t}) \label{eq3-27-0}\\
&=N_{s_t}\left\lfloor \frac{x_{t-1}}{N}\right\rfloor+\bmod(x_{t}, N_{s_t}) \nonumber\\
&=N_{s_t}\left\lfloor \frac{x_{t-1}}{N}\right\rfloor+\bmod(x_{t-1},N)-d_{s_t}. \nonumber
\end{align}

Figure \ref{fig-rANS} shows the relations of  $x_t$ and $x_{t-1}$ on the number line, which are given by \eqref{eq3-21}, \eqref{eq3-26}--\eqref{eq3-27-0}.

\vspace{0.3cm}
\subsection{Average code length of rANS} \label{sec2.1}
We evaluate the average code length of rANS based on the source probability distribution
$p=\{p(s)\;|\; s\in\mathcal{S}\}$ and the probability distribution $q=\{q(s)\;|\; s\in\mathcal{S}\}$,
which is defined by $q(s)=N_{s}/N$ for $s\in\mathcal{S}$.

From \eqref{eq3-26}, we have
\begin{align}
 \frac{x_{t-1}}{N}&\approx \frac{x_{t}}{N_{s_t}}, \nonumber\\
\frac{x_{t-1}}{x_t}&\approx \frac{N}{N_{s_t}} =\frac{1}{q(s_t)}.\label{eq3-29}
\end{align}
We first consider the case such that \eqref{eq3-29} holds with ``$=$'' instead of ``$\approx$''.
Let $x_0(s^T)$ denote the codeword of $s^T=s_1s_2\cdots s_T$ obtained by applying  \eqref{eq3-21} to $s_t$, $t=T, T-1, \cdots, 1$, for $x_T=1$ in this case.
Then, the bit length of $x_0(s^T)$ is given by
\begin{align}
\lg x_0(s^T)&=\lg \prod_{t=1}^T \frac{x_{t-1}}{x_t}+\lg x_T\nonumber\\
&=\sum_{t=1}^T \lg\frac{x_{t-1}}{x_t}\nonumber\\
&= \sum_{t=1}^T \lg \frac{1}{q(s_t)}. \nonumber
\end{align}
Hence, the average code length $L$ can be expressed by
\begin{align}
L&=\frac{1}{T}\sum_{s^T\in\mathcal{S}^T}p(s^T)\lg x_0(s^T) \nonumber\\
&=\frac{1}{T}\sum_{s^T\in\mathcal{S}^T}p(s^T)\sum_{t=1}^T\lg \frac{1}{q(s_t)} \nonumber\\
&=\frac{1}{T}\sum_{t=1}^T\sum_{s_t\in\mathcal{S}}p(s_t)\lg \frac{1}{q(s_t)} \nonumber\\
&=\sum_{s\in\mathcal{S}} p(s)\lg \frac{1}{q(s)} \hspace{0.7cm} \nonumber\\
&= H(p)+D(p\|q), \label{eq3-31}
\end{align}
where $H(p)$ and $D(p\|q)$ are the source entropy and the relative entropy to $q$, respectively.

We next estimate the degree of approximation of \eqref{eq3-29}.
Since it holds from \eqref{eq3-26} that
\begin{align}
\frac{x_{t-1}}{N}-1< \frac{x_{t}}{N_{s_t}}, \quad \frac{x_{t-1}}{N}> \frac{x_{t}}{N_{s_t}} -1, \nonumber
\end{align}
$x_{t-1}/x_t$ has the following upper and lower bounds:
\begin{align}
\frac{N}{N_{s_t}}-\frac{N}{x_t}<\frac{x_{t-1}}{x_t}<\frac{N}{N_{s_t}}+\frac{N}{x_t}. \label{eq3-34}
\end{align}
These bounds mean that \eqref{eq3-29} is a very good approximation when $x_t$
satisfies
\begin{align}
x_t\gg N. \label{eq3-35}
\end{align}

We now evaluate the influence of approximation errors on the code length.
From the right inequality of \eqref{eq3-34}, we have
\begin{align}
\lg \frac{x_{t-1}}{x_t}&<\lg \left(\frac{N}{N_{s_t}}+\frac{N}{x_t}\right) \nonumber\\
&=\lg\frac{N}{N_{s_t}}+\lg \left(1+\frac{N_{s_t}}{x_t}\right)\label{eq3-37-0}\\
&<\lg \frac{1}{q(s_t)} +\lg \left(1+\frac{N}{x_{t}}\right) \nonumber\\
&<  \lg \frac{1}{q(s_t)} +(\lg e)\frac{N}{x_{t}}. \label{eq3-37}
\end{align}
The second term of \eqref{eq3-37} represents an upper bound on code length loss 
due to the approximation error in \eqref{eq3-29}.
Hence the total loss $l_{\text{loss}}(s^T)$ of the data sequence $s^T$ is bounded by
\begin{align}
l_{\text{loss}}(s^T)<(N\lg e) \sum_{t=1}^T \frac{1}{x_t}. \nonumber
\end{align}

Consider the case of $x_T=A$, where $A$ is an integer satisfying  
 $\displaystyle{\eta=\left(\min_{s\in\mathcal{S}}\frac{N}{N_s}\right)-\frac{N}{A}>1}$.
 Then from the left inequality of \eqref{eq3-34} and $A = x_T < x_{T-1} <\cdots< x_{t-1}$,
 we obtain $x_{t-1}>\eta^{T-t+1}A$.
 In this case, adding the loss $\lg A$ caused by using $x_T=A$ instead of $x_T=1$,
 the total loss $l_{\text{loss}}(s^T)$ has the following upper bound.
\begin{align}
l_{\text{loss}}(s^T)
&<\lg A + (N\lg e) \sum_{t=1}^T \frac{1}{\eta^{T-t}A}\nonumber\\
&<\lg A + \frac{N\lg e}{A} \frac{\eta}{\eta-1}. \label{eq3-39}
\end{align}

Therefore, from \eqref{eq3-31} and \eqref{eq3-39},
the average code length $L$ in the case of $x_T=A$ satisfies
\begin{align}
L&<H(p)+D(p\|q)+\frac{1}{T} \left(\lg A + \frac{N\lg e}{A} \frac{\eta}{\eta-1}\right).\label{eq3-39-2}
\end{align}
We note from \eqref{eq3-39-2} that $L\to H(p)+D(p\|q)$ as $T\to\infty$.

\section{rANS with a finite digit arithmetic}
The rANS treated  in the previous section has a practical drawback for large $T$ because 
$x_t$ becomes a very large integer as $t$ approaches zero.
To overcome this drawback, the stream coding is designed so that
encoding and decoding can be performed with fixed finite digits, no matter how large $T$ is.

In this section, we explain Townsend's encoding and decoding algorithms \cite{Townsend},
which are a stream version of the rANS, and evaluate its average code length. 
We assume in this section that $N$ and $N_s$ satisfy $N=\sum_{s\in\mathcal{S}}N_s=2^R$
for a given integer $R$.

\begin{figure*}[t]
 \begin{center}
   \includegraphics[width=15cm]{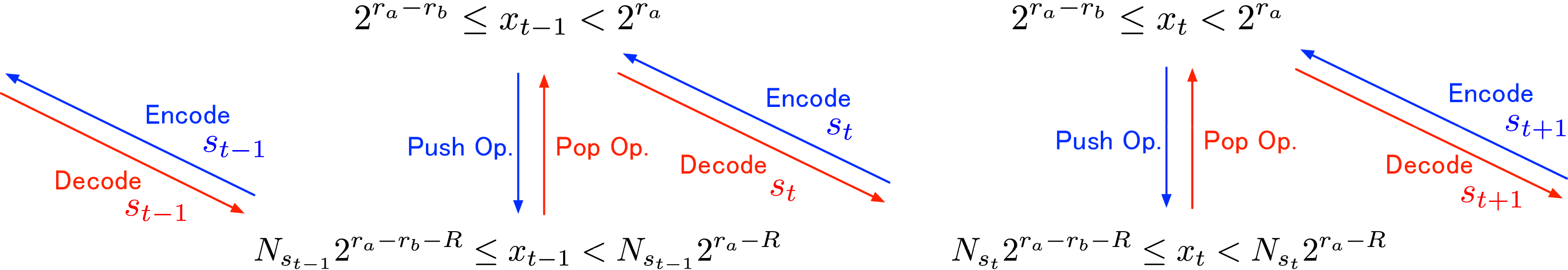}
    \caption{Relation between Push and Pop used in stream encoding and decoding of rANS.}
      \label{fig-PushPop}
 \end{center}
\end{figure*}

\subsection{Encoding and decoding procedures of rANS using finite digit operations \cite{duda2014}}
From \eqref{eq3-21}--\eqref{eq3-23}, the encoding and decoding functions are
defined in the case of $N=2^R$ as follows.

\vspace{0.1cm}
\noindent \textbf{A. Definition of encoding function}\\
Encoding function $x_{t-1}:=C(s_t, x_{t})$ is defined by
\begin{align}
x_{t-1}&:=2^R\left\lfloor \frac{x_t}{N_{s_t}}\right\rfloor +d_{s_t} + \bmod(x_t, N_{s_t}).\label{eq4-49}
\end{align}

\vspace{0.1cm}
\noindent \textbf{B. Definition of decoding function}\\
Decoding function $(s_t, x_t):=D(x_{t-1})$ is defined by
\begin{align}
s_t&:=\min\left\{s: \bmod(x_{t-1},2^R)<\sum_{i\preceq s} N_i\right\}, \nonumber\\
x_t&:= N_{s_t}\left\lfloor \frac{x_{t-1}}{2^R}\right\rfloor+\bmod(x_{t-1},2^R)-d_{s_t}.\nonumber
\end{align}

We now add the following restriction to $x_{t-1}$ obtained by \eqref{eq4-49}.
\begin{align}
2^{r_a-r_b}\leq x_{t-1}< 2^{r_a},  \label{eq4-44}
\end{align}
where $r_a$ and $r_b$ are integers satisfying $r_a-r_b>R$.
Then, from \eqref{eq4-49} and \eqref{eq4-44}, we have the following inequalities.
\begin{align}
2^{r_a-r_b}&\leq 2^R\left\lfloor \frac{x_{t}}{N_{s_t}}\right\rfloor +d_{s_t} + \bmod(x_t, N_{s_t}) <2^{r_a}, \nonumber\\
2^{r_a-r_b-R}&\leq \left\lfloor \frac{x_{t}}{N_{s_t}}\right\rfloor +\frac{d_{s_t} + \bmod(x_t, N_{s_t})}{2^R}<2^{r_a-R},\nonumber\\
2^{r_a-r_b-R}&\leq \left\lfloor \frac{x_{t}}{N_{s_t}}\right\rfloor <2^{r_a-R},\label{eq4-45-0}\\
2^{r_a-r_b-R}&\leq \frac{x_{t}}{N_{s_t}} <2^{r_a-R},\nonumber\\
N_{s_t}2^{r_a-r_b-R}&\leq x_{t} <N_{s_t}2^{r_a-R},  \label{eq4-45}
\end{align}
where \eqref{eq4-45-0} holds because, from \eqref{eq3-24}, we have $(d_{s_t} + \bmod(x_t, N_{s_t}))/2^R <1$.

In order for $x_{t-1}$ obtained by \eqref{eq4-49} to satisfy \eqref{eq4-44},
$x_{t}$ must satisfy $\eqref{eq4-45}$.
Therefore, if $x_{t}$ does not satisfy \eqref{eq4-45}, 
we remove the lower bits of $x_{t}$ after pushing them onto a stack, and then
we perform the encoding of \eqref{eq4-49}.
On the other hand, in decoding, $x_t$ decoded from $x_{t-1}$ satisfying \eqref{eq4-44} satisfies
\eqref{eq4-45}. Therefore, by popping the lower bits of $x_{t}$ from the stack and adding them to $x_{t}$,
$x_{t}$ can satisfy $2^{r_a-r_b}\leq x_{t}< 2^{r_a}$. After finishing this process, we move on to the step of
decoding $s_{t+1}$.

Figure \ref{fig-PushPop} shows the relation between Push and Pop operations in encoding and decoding.
Let $\text{stack-push }(u, \tilde{u})$ stand for pushing $\tilde{u}$ into a stack $u$, and 
let $\text{stack-pop }(u)$ represent popping $u_{\text{top}}$ with an appropriate length from
the top of the stack $u$.
Then, the Push and Pop operations are described as follows.

\vspace{0.2cm}
\noindent\textbf{Push operation for $\bm{x_t}$}
\vspace{-0.3cm}
\begin{align}
\text{\textbf{While} } x_{t}\geq& N_{s_t}2^{r_a-R}: \nonumber\\
 u&\leftarrow \text{stack-push }(u, \bmod(x_{t}, 2^{r_b})),  \label{eq4-46}\\
 x_{t}&\leftarrow\left\lfloor\frac{x_{t}}{2^{r_b}}\right\rfloor.\nonumber
\end{align}

\vspace{0.2cm}
\noindent\textbf{Pop operation for $\bm{x_t}$}
\vspace{-0.3cm}
\begin{align}
\text{\textbf{While} } x_{t}<& 2^{r_a-r_b}: \nonumber\\
 u_{\text{top}}&\leftarrow \text{stack-pop }(u), \hspace*{1.2cm}\label{eq4-47}\\
 x_{t}&\leftarrow 2^{r_b}x_t+u_{\text{top}}.\hspace*{1.2cm}\nonumber
\end{align}

In the above Push operation, the number of times that Eq.~\eqref{eq4-46} is applied depends on
the values of $s_t$ and $x_t$. If $x_t$ satisfying $2^{r_a-r_b}\leq x_t< 2^{r_b}$
also satisfies $x_t <N_{s_t}2^{r_a-R}$ (i.e., if $2^{-r_b}<p(s_t)\approx q(s_t)=N_{s_t}/N=N_{s_t}2^{-R}$
and $2^{r_a-r_b}\leq x_t< N_{s_t}2^{r_a-R}$), Eq.~\eqref{eq4-46} is not executed. 
On the other hand, if $2^{-r_b}>p(s_t)\approx q(s_t)=N_{s_t}/N=N_{s_t}2^{-R}$ and 
$N_{s_t}2^{r_a-R}\leq x_t 2^{-r_b}< 2^{r_a-r_b}$, then Eq.~\eqref{eq4-46} is executed more than once.
We note that in order to satisfy \eqref{eq3-35},  parameters $r_a, r_b$, and $R$ must satisfy
\begin{align}
2^{r_a-r_b}\gg 2^R. \label{eq4-48}
\end{align}

Combining Pop-Push operations with encoding and decoding functions,
we obtain the following encoding and decoding algorithms.

\vspace{0.2cm}
\noindent \textbf{C. Encoding algorithm}
\vspace{-0.2cm}
\begin{enumerate}[a.]
 \setlength{\parskip}{0cm}
 \setlength{\itemsep}{0cm}
 \setlength{\itemindent}{0cm}
 \setlength{\labelsep}{0.2cm}
\item For a given data sequence $s^T=s_1s_2\cdots s_T$, set $x_T\leftarrow 2^{r_a-r_b}$.
\item Repeat the following (i) and (ii) in backward order, $t=T, \cdots, 2, 1$.\\
\hspace*{0.5cm}(i)\; Perform the Push operation on $x_t$,\\
\hspace*{0.5cm}(ii) $x_{t-1}\leftarrow C(s_t, x_{t})$.
\item  The codeword of $s^T$ is given by $x_0$ and stack $u$.
\end{enumerate}

\vspace{0.2cm}
\noindent \textbf{D. Decoding algorithm}
\vspace{-0.2cm}
\begin{enumerate}[a.]
 \setlength{\parskip}{0cm}
 \setlength{\itemsep}{0cm}
 \setlength{\itemindent}{0cm}
 \setlength{\labelsep}{0.2cm}
\item Set a codeword ($x_0$ and stack $u$).
\item Repeat the following (iii) and (iv) in forward order, $t=1, 2, \cdots, T$.\\
\hspace*{0.4cm}(iii) $(s_t, x_t)\leftarrow D(x_{t-1})$, \\
\hspace*{0.4cm}(iv)\; Perform the Pop operation on $x_t$.
\item $s^T=s_1s_2\cdots s_T$ is the decoded sequence.
\end{enumerate}

\begin{remark}{\rm 
If a similar method is applied to the ABS in Section 2, it can be encoded and decoded using finite digit operations for any data sequence.
}
\end{remark}

\subsection{Average code length of stream rANS}
When we encode $s_t$ at time $t$, the increase in code length is given by $l(s_t)=\lg x_{t-1}-\lg x_t=\lg (x_{t-1}/x_t)$. 
Hence, combining \eqref{eq3-37-0} with the left inequality in \eqref{eq4-45},
we can derive the following upper bound of $l(s_t)$.
\begin{align}
l(s_t)
=\lg \frac{x_{t-1}}{x_t}&< \lg\frac{N}{N_{s_t}}+\lg \left(1+\frac{N_{s_t}}{x_t}\right)\nonumber\\
&\leq \lg\frac{N}{N_{s_t}}+\lg\left(1+\frac{1}{2^{r_a-r_b-R}}\right)\nonumber\\
&< \lg\frac{1}{q(s_t)}+\frac{\lg \text{e}}{2^{r_a-r_b-R}}.\nonumber
%
\end{align}
Hence, in the case of $x_T=2^{r_a-r_b}$,
the total code length $l(s^T)$ is given by
\begin{align}
l(s^T)
&=\sum_{t=1}^T l(s_t)+ \lg x_T\nonumber\\
&<\sum_{t=1}^T \left(\lg\frac{1}{q(s_t)}+\frac{\lg \text{e}}{2^{r_a-r_b-R}} \right)+r_a-r_b\nonumber\\
&= \sum_{t=1}^T\lg\frac{1}{q(s_t)}+ \left(\frac{T\lg \text{e}}{2^{r_a-r_b-R}}+r_a-r_b\right).\label{eq4-53}
\end{align}

In the same way as \eqref{eq3-31}, 
the average code length $L=(1/T)\sum p(s^T)l(s^T)$ can be bounded from \eqref{eq4-53} as follows.
\begin{align}
L<  H(p)+D(p\|q)+\frac{\lg \text{e}}{2^{r_a-r_b-R}} +\frac{r_a-r_b}{T},\label{eq4-54}
\end{align}
where the 3rd and 4th terms on the right side  become
sufficiently small when we use parameters $r_a, r_b, R$ satisfying \eqref{eq4-48} and $T\gg r_a-r_b$.

\section{tANS}

In this section, we discuss the tabled variant of ANS (tANS), which is sometimes simply referred to as ANS \cite{PDPCMM}\cite{PDPCMM2023}.
For any $|\mathcal{S}|\geq 2$, 
tANS can encode and decode data sequences using only integer operations like rANS.
In the case of the stream rANS, a data sequence $s^T$ is encoded to ($x_0$ and stack $u$),
but a codeword is not determined for each $s_t$.
On the other hand, the tANS is designed so that the codeword of $s_t$ is uniquely determined 
from $s_t$ and $x_t$.
As a result, tANS has the advantage that encoding and decoding can be performed entirely using tables without arithmetic operations, as will be described later.

In the same way as the previous section, we assume that $N$ and $N_s, s\in\mathcal{S}$
satisfy $N=\sum_{s\in\mathcal{S}}N_s=2^R$ for a given integer $R$.
Then the encoding and decoding of tANS are defined by using sets of integer states $\mathcal{X}$, $\mathcal{X}_s$, and $\mathcal{Y}$.

\vspace{-0.2cm}
\begin{enumerate}[1.]
 \setlength{\parskip}{0cm}
 \setlength{\itemsep}{0cm}
 \setlength{\itemindent}{0.cm}
 \setlength{\labelsep}{0.2cm}
\item $\mathcal{X}$: The set of states used in encoding and decoding, which is defined as $\mathcal{X}=\{N, N+1, \cdots, 2N-1\}$ and $N=|\mathcal{X}|$.
\item $\mathcal{X}_s$: The set of states corresponding to $s\in \mathcal{S}$, which satisfies
$\mathcal{X}_s\cap \mathcal{X}_{s'}=\emptyset$ for $s\neq s'$, and $\mathcal{X}=\bigcup_{s\in\mathcal{S}}\mathcal{X}_s$. Then,  $N=\sum_{s\in\mathcal{S}}N_s$ for $N_s=|\mathcal{X}_s|$.
\item $\mathcal{Y}_s$:  The other set of states corresponding to $s\in \mathcal{S}$, which 
is defined as $\mathcal{Y}_s=\{N_s, N_s+1, \cdots, 2N_s-1\}$ and $N_s=|\mathcal{Y}_s|$.
\end{enumerate}

Since it holds that $N_s=|\mathcal{X}_s|=|\mathcal{Y}_s|$ for each $s\in\mathcal{S}$,  we have a one-to-one correspondence between $x\in\mathcal{X}_s$ and $y\in\mathcal{Y}_s$.
Furthermore, since $\mathcal{X}_s$ also satisfies the above condition 2, 
we also have a one-to-one correspondence between a pair $(s,y)$, $s\in\mathcal{S}$, $y\in\mathcal{Y}_s$, and $x\in\mathcal{X}$.
We represent this one-to-one correspondence by encoding and decoding functions $\tilde{C}$ and $\tilde{D}$.
\footnote{We use notation $(\tilde{C}, \tilde{D})$ since these functions have distinct meanings from the encoding function $x_{t-1}:=C(s_t, x_t)$ and the decoding function $(s_t, x_t):=D(x_{t-1})$ used in previous sections.}

\vspace{0.2cm}
\noindent \textbf{A. Definition of encoding function}\\
For each $s\in\mathcal{S}$, encoding function $\tilde{C}(s, \;\cdot\;)$ is a bijection function $\tilde{C}(s, \;\cdot\;): \mathcal{Y}_s\to\mathcal{X}_s$.
Note that if $s\in\mathcal{S}$ and $y\in\mathcal{Y}_s$, then $x:=\tilde{C}(s,y)\in\mathcal{X}_s$.

\vspace{0.2cm}
\noindent \textbf{B. Definition of decoding function}\\
For each $x\in\mathcal{X}_s$, decoding function $\tilde{D}$ is a bijection function $\tilde{D}: \mathcal{X}_s \to \{s\}\times \mathcal{Y}_s$.
Note that if $x\in\mathcal{X}_s\subset \mathcal{X}$, then $(s,y):=\tilde{D}(x)\in \{s\}\times \mathcal{Y}_s$.

\vspace{0.2cm}
\subsection{Encoding and decoding procedures of tANS \cite{duda2014}\cite{PDPCMM}\cite{PDPCMM2023}}

Let $x_t, s_t, b_t$, and $k_t$ represent the state, data symbol, codeword, and code length at time $t$,
respectively.
Then,  encoding is the process of obtaining $(b_t, x_{t-1})$ from $(x_t, s_t)$, while
decoding is the process of obtaining $(s_t, x_t)$ from $(x_{t-1}, b_t$).
In the tANS,  these processes are performed via $(y_{t-1}, k_t)$ as shown in 
the following encoding and decoding algorithms \cite{duda2014}\cite{PDPCMM},
where it is assumed that functions $\tilde{C}$ and $\tilde{D}$ are given.

\vspace{0.2cm}
\noindent \textbf{C. Encoding algorithm}
\vspace{-0.2cm}
\begin{enumerate}[a.]
 \setlength{\parskip}{0cm}
 \setlength{\itemsep}{0cm}
 \setlength{\itemindent}{0cm}
 \setlength{\labelsep}{0.2cm}
\item For a given data sequence $s^T=s_1s_2\cdots s_T$, select $x_T\in\mathcal{X}$ arbitrarily.
\item Repeat the following calculations in backward order, $t=T, \cdots, 2, 1$.
\begin{align}
k_t&\leftarrow \left\lfloor \lg \frac{x_t}{N_{s_t}}\right\rfloor, \label{eq5-1}\\
b_t&\leftarrow \bmod (x_t, 2^{k_t}), \label{eq5-2}\\
y_{t-1}&\leftarrow\left\lfloor \frac{x_t}{2^{k_t}}\right\rfloor, \label{eq5-3}\\
x_{t-1}&\leftarrow\tilde{C}(s_t, y_{t-1}). \label{eq5-4}
\end{align}
\item
The codeword sequence of $s^T$ is $x_0b_1b_2\cdots b_T$.
\end{enumerate}

\vspace{0.2cm}
\noindent \textbf{D. Decoding algorithm}
\vspace{-0.2cm}
\begin{enumerate}[a.]
 \setlength{\parskip}{0cm}
 \setlength{\itemsep}{0cm}
 \setlength{\itemindent}{0cm}
 \setlength{\labelsep}{0.2cm}
\item For a given codeword sequence $x_0b_1b_2\cdots b_T$, set $x_0$ and $\bm{b}\leftarrow b_1b_2$ $\cdots b_T$.
\item Repeat the following calculations in forward order, $t=1, 2, \cdots, T$.
\vspace{-0.2cm}
\begin{align}
\hspace*{-0.5cm}(s_t, y_{t-1})&\leftarrow\tilde{D}(x_{t-1}), \nonumber\\
k_t&\leftarrow R-\lfloor\lg y_{t-1}\rfloor, \label{eq5-6}\\
b_t&\leftarrow\text{the fitst $k_t$ bits of $\bm{b}$}, \label{eq5-6-2}\\
\bm{b}&\leftarrow\text{the sequence obtained by removing $b_t$ from $\bm{b}$},\nonumber\\
x_t&\leftarrow 2^{k_t} y_{t-1} +b_t. \label{eq5-7}
\end{align}
\item $s^T=s_1s_2\cdots s_T$ is the decoded sequence.
\end{enumerate}
\vspace{0.2cm}

State $x_0$ included in the codeword sequence can be represented with, e.g., $\lg N=R$ bits if we use the fixed length code. From \eqref{eq5-2}, we note that $k_t$ stands for the bit length of $b_t$.
In decoding, since $k_t$ can be obtained by \eqref{eq5-6}, we can extract $b_t$ from the
codeword sequence $\bm{b}$ in \eqref{eq5-6-2}.

\begin{remark}{\rm
Encoding and decoding algorithms include the calculation of $\lfloor \lg a\rfloor$ for 
a positive integer $a$.
But, since $\lfloor \lg a\rfloor$ is the number of bits of $a$ excluding the MSB,  it can  easily be obtained without logarithm calculation.
}\end{remark}

\begin{remark}{\rm
Eq.~\eqref{eq5-2} corresponds to the operation of \eqref{eq4-46}.
In Eq.~\eqref{eq4-46}, the remainder is taken by $2^{r_b}$ regardless of the values of $s_t$ and $x_t$, whereas in Eq.~\eqref{eq5-2}, the remainder is taken by $2^{k_t}$, which is determined 
by the values of $s_t$ and $x_t$, so that the code word $b_t$ is determined for each $s_t$.
}\end{remark}

We first show that $y_{t-1}$ given by \eqref{eq5-3} satisfies $y_{t-1}\in\mathcal{Y}_{s_t}$, i.e.,
$N_{s_t}\leq y_{t-1}<2N_{s_t}$.
From \eqref{eq5-1}, we have the following inequalities.
\begin{align}
2^{k_t}&\leq \frac{x_t}{N_{s_t}} <2^{k_t+1},\nonumber\\
N_{s_t}&\leq \frac{x_t}{2^{k_t}}< 2 N_{s_t},\nonumber\\
N_{s_t}&\leq y_{t-1}=\left\lfloor \frac{x_t}{2^{k_t}}\right\rfloor< 2 N_{s_t}, \label{eq5-11-5}
\end{align}
where \eqref{eq5-11-5} holds because $N_{s_t}$ is an integer.
Since \eqref{eq5-11-5} means $y_{t-1}\in\mathcal{Y}_{s_t}$, we can use the encoding function
$\tilde{C}(s_t, y_{t-1})$ in \eqref{eq5-4} to obtain $x_{t-1}=\tilde{C}(s_t, y_{t-1})\in\mathcal{X}_{s_t}$.

Since $b_t$ and $y_{t-1}$ are given by \eqref{eq5-2} and \eqref{eq5-3}, respectively, in encoding,
$x_t$ can be decoded by \eqref{eq5-7} in decoding.
Next we show that $k_t$ can be obtained by \eqref{eq5-6}.
From $x_t\in\mathcal{X}$ (i.e., $N\leq x_t< 2N$), \eqref{eq5-7}, and $N=2^R$,
we have the following inequalities.
\begin{align}
2^R&\leq 2^{k_t} y_{t-1} +b_t < 2^{R+1}, \nonumber\\
2^{R-k_t}&\leq y_{t-1}+\frac{b_t}{2^{k_t}} <2^{R-k_t+1}, \nonumber\\
2^{R-k_t}&\leq y_{t-1} <2^{R-k_t+1}, \label{eq5-14}\\
R-k_t&\leq \lg y_{t-1} < R-k_t+1, \nonumber
\end{align}
which means that $k_t$ is obtained by \eqref{eq5-6}. Note that \eqref{eq5-14} holds
because $2^{R-k_t}$ and $y_{t-1}$ are integers and we have $0\leq b_t/2^{k_t}<1$ from \eqref{eq5-2}.

\begin{remark}{\rm 
Although we consider the case of $N=2^R$ in the above, tANS can also be applied to general cases 
where $N$ is not a power of 2 \cite{duda2014}.
But, in decoding of general cases, $k_t$ is given by $k_t=\lceil \lg (N/y_{t-1})\rceil$ or $k_t=\lceil \lg (N/y_{t-1})\rceil-1$ instead of \eqref{eq5-6},
and it is necessary to use $k_t$ that satisfies $N\leq 2^{k_t}y_{t-1}+b_t<2N$.
}\end{remark}

\begin{remark}{\rm 
If we precalculate Eqs.~\eqref{eq5-1}--\eqref{eq5-4} and \eqref{eq5-6}--\eqref{eq5-7} for all cases of $x_t=x, N\leq x<2N$ and $s_t=s, s\in\mathcal{S}$ and store the results as a table, we can perform encoding and decoding without performing arithmetic calculations each time.
For this reason, it is called tANS (tabled-variant of ANS).
}\end{remark}

\subsection{Average code length of tANS}

The influence of the choice of $x_T\in\mathcal{X}$ in encoding and the increase in the average code length due to the bit length $R$ of $x_0$ contained in the codeword converge to zero as the data sequence length $T$ becomes longer. For simplicity, in this section we will ignore these and find the average code length in the steady state.

Let $Q(x)$ denote the stationary probability of $x\in\mathcal{X}$, and let $q(s)$ be defined by $q(s)=N_s/N$.
We first evaluate the average code length $L$ based on \eqref{eq5-1} as follows.
\begin{align}
L&=\sum_{s\in\mathcal{S}}\sum_{x\in\mathcal{X}}p(s)Q(x) \left\lfloor\lg \frac{x}{N_{s}}
\right\rfloor \nonumber\\
&\leq \sum_{s\in\mathcal{S}}\sum_{x\in\mathcal{X}}p(s)Q(x)\lg \frac{x}{N_{s}}
\nonumber\\
&=\sum_{s\in\mathcal{S}}\sum_{x\in\mathcal{X}}p(s)Q(x)\left(\lg \frac{N}{N_{s}}+\lg \frac{x}{N}\right)\nonumber\\
&=\sum_{s\in\mathcal{S}}p(s)\lg \frac{N}{N_{s}}+\sum_{x\in\mathcal{X}}Q(x)\lg \frac{x}{N}
\label{eq5-20-0}\\
&\leq^{*1} \sum_{s\in\mathcal{S}}p(s)\lg \frac{1}{q(s)}+\lg \frac{\sum_{x\in\mathcal{X}}Q(x)x}{N}\nonumber\\
&= H(p)+D(p\|q) + \lg \frac{\text{E}[X]}{N},  \label{eq5-20}
\end{align}
where $\leq^{*1}$ comes from Jensen's inequality for the $\lg$ function, and
$\text{E}[X]$ is the expected value of $x$ in the steady state.
Since $N\leq x\leq 2N-1$, a loose upper bound  $\lg (\text{E}[X]/N)<1$ holds.

Unlike ABS and rANS treated in previous sections, tANS encodes each $s_t$ symbol-by-symbol to a codeword with $k_t$ bits. Therefore, there is a loss in the average code length 
compared to assigning one codeword at a time to the entire $s^T=s_1s_2\cdots s_T$. 
Below, we evaluate this loss under the tANS conditions, i.e., $N\leq x_{t-1}<2N$ and $N_{s_t}\leq y_{t-1}<2N_{s_t}$.

If we use $\tilde{x}_{t-1}:=(N/N_{s_t})y_{t-1}$ instead of $x_{t-1}$,
we can satisfy
\begin{align}
N\leq \tilde{x}_{t-1}=\frac{N}{N_{s_t}}y_{t-1}< 2N. \label{eq5-60}
\end{align}
However, $\tilde{x}_{t-1}$ is generally not an integer. Therefore,
tANS uses a function $\tilde{C}$ to map $y_{t-1}$ to an integer $x_{t-1}$ that satisfies $N\leq x_{t-1}<2N$.

Corresponding to the case where $s^T$ is encoded all at once, 
we consider the ideal case where real-valued codeword length is allowed for each encoding of $s_t$.
The codeword length in this case is given by $l(s_t)=\lg(x_{t-1}/y_{t-1})$, as in the case of rANS.

If $\tilde{x}_{t-1}$ is used, then, from \eqref{eq5-60},  $l(s_t)=\lg(\tilde{x}_{t-1}/y_{t-1})=\lg (N/N_{s_t})=\lg 1/q(s_t)$, which depends only on the value of $s_t$ and not on the value of $y_{t-1}$.
Therefore, the average code length $\tilde{L}^*$ of this case is given by
\begin{align}
\tilde{L}^*&=\sum_{s\in\mathcal{S}} p(s)\lg \frac{N}{N_{s}}\nonumber\\
&=H(p)+D(p\|q). \nonumber
\end{align}
However, when we use $x_{t-1}=\tilde{C}(s, y_{t-1})$, 
the value of $x_{t-1}/y_{t-1}$ deviates slightly from $N/N_{s_t}$,
so the average code length becomes longer than $\tilde{L}^*$.
The real-valued codeword length for $x_{t-1}=\tilde{C}(s_t, y_{t-1})$ is 
given by $l(s_t)=\lg (\tilde{C}(s_t, y_{t-1})/y_{t-1})$, which depends on
both $s_t$ and $y_{t-1}$, so the average code length $L^*$ can be evaluated as follows.
\begin{align}
L^*&=\sum_{s\in\mathcal{S}} \sum_{y\in\mathcal{Y}_s} Q(\tilde{C}(s,y)) \lg \frac{\tilde{C}(s,y)}{y} \nonumber\\
&=\sum_{s\in\mathcal{S}} \sum_{y\in\mathcal{Y}_s} Q(\tilde{C}(s,y)) \left(\lg \frac{N}{N_s}+\lg \frac{\tilde{C}(s,y)N_s}{Ny}  \right)\nonumber\\
&=^{*2}  \sum_{s\in\mathcal{S}} p(s) \lg \frac{N}{N_s}
+\sum_{s\in\mathcal{S}} \sum_{y\in\mathcal{Y}_s} Q(\tilde{C}(s,y))\lg \frac{\tilde{C}(s,y)}{N} 
-\sum_{s\in\mathcal{S}} \sum_{y\in\mathcal{Y}_s} Q(\tilde{C}(s,y))\lg \frac{y}{N_s}\nonumber\\
 &=^{*3}  \sum_{s\in\mathcal{S}} p(s) \lg \frac{N}{N_s}+\sum_{x\in\mathcal{X}}Q(x)\lg \frac{x}{N} 
-\sum_{s\in\mathcal{S}} \sum_{y\in\mathcal{Y}_s} Q(\tilde{C}(s,y))\lg \frac{y}{N_s},\label{eq5-62}
\end{align}
where the numbered equalities hold because
\begin{description}
\item{$=^{*2}$:} $\sum_{y\in \mathcal{Y}_s} Q(\tilde{C}(s,y))=p(s)$,
\item{$=^{*3}$:} $(s,y)$ and $x=\tilde{C}(s,y)$ have a one-to-one correppondence.
\end{description}

From \eqref{eq5-20-0} and \eqref{eq5-62}, the loss of average code length of tANS compared 
with $L^*$ is bounded by 
\begin{align}
L-L^*
&\leq \sum_{s\in\mathcal{S}} \sum_{y\in\mathcal{Y}_s} Q(\tilde{C}(s,y))\lg \frac{y}{N_s}\nonumber\\
&\leq^{*1} \sum_{s\in\mathcal{S}} p(s) \lg\frac{\sum_{y\in\mathcal{Y}_s} \frac{Q(\tilde{C}(s,y))}{p(s)} y}{N_s}\nonumber\\
&=\sum_{s\in\mathcal{S}} p(s)\lg \frac{\text{E}[Y|s]}{N_s}, \label{eq5-65}
\end{align}
where $\text{E}[Y|s]$ is the expected value of $y_{t-1}$ under the condition of $s_t=s\in\mathcal{S}$. Since we have $\text{E}[Y|s]/N_s<2$ in \eqref{eq5-65} from \eqref{eq5-11-5},
a looser bound is given by $L-L^*<1$.

Although \eqref{eq5-20} is an upper bound for any $\tilde{C}$,
the average code length $L$ can be made smaller by choosing $\tilde{C}$ appropriately.
Methods of constructing $\tilde{C}$ with good performance have been proposed in references such as \cite{PDPCMM} \cite{PDPCMM2023} \cite{Y2023}.
It is also shown in \cite{Yokoo2016} \cite{DY2019} \cite{YD2019} that
the ideal stationary probability distribution\footnote{If $Q^*(x)$ can be realized, it is optimal.} of tANS is given by
\begin{align}
Q^*(x)=\lg \frac{x+1}{x},\nonumber
\end{align}
and paper \cite{Y2023} shows how to construct $\tilde{C}$ based on $Q^*(x)$.

Furthermore, it is proved in  \cite{YI-sita2023} that if $Q(x)$ satisfies $Q(x)\leq Q^*(x)+(\alpha/N^2)$ for every $x\in\mathcal{X}$ and a constant $\alpha\geq 0$, the average code length $L$ is bounded by
\begin{align}
L\leq H(p)+D(p\|q) +\frac{\alpha}{N}. \nonumber\label{eq5-71}
\end{align}
In the case of $q(s)=N/N_s=p(s)$, a construction method of $\tilde{C}$ is shown in 
\cite{YD2019}\cite{YI-sita2023} such that $Q(x)$ satisfies $Q(x)= Q^*(x)+O(1/N^2)$.

\section{Conclusions}
In this paper, we explained in detail the encoding and decoding algorithms for variants of ANS,  
and we derived several upper bounds on their average coding lengths.

We assumed that source probability distribution $p=\{p(s)\}$ is given. But, if $p$ is unknown,
it is necessary to include information on $p$ or $\{N_s\}$ in codewords.
Alternatively, it is possible to use a frequency distribution like dynamic Huffman code \cite{Gallager},
without including information on $p$ or $\{N_s\}$ in codewords.
When decoding $s_{t+1}$ using the probability distribution $\hat{p}_t=\{\hat{p}_t(s)\}$ estimated based on the frequency distribution of $s_1^t=s_1s_2\cdots s_t$, the encoding also requires that the probability distribution $\hat{p}_t, t=1,2\cdots, T$ is first obtained from $s^T$, and then encoding is performed using $\hat{p}_t$ in the backward order, that is, $t=T, \cdots, 2, 1$.
In this paper, we have explained that $s^T$ is encoded in backward order ($s_t, t=T, T-1,\cdots, 1$)
and decoded in forward order ($s_t, t=T, T-1,\cdots, 1$). 
But, it is also possible to perform encoding in forward order and decoding in backward order.

The ANS is designed to perform encoding and decoding using integer arithmetic operations, similar to 
Range code of arithmetic cording. 
However, even without using arithmetic operations, 
high-performance compression is possible by performing encoding and decoding in reverse order \cite{YI-IT2023}--\cite{YI-sita2023}.
There has also been research into using the ANS as a simple cipher or a simple random number generator for cryptography \cite{DN2016}--\cite{CDMMNPP2022}\cite{PPMMDC}.


\section*{Acknowledgement}
This work was supported by JSPS KAKENHI Grant Number 24K07487.





\end{document}